\documentclass[pra,twocolumn,amsmath,amssymb,10pt,aps,superscriptaddress,longbibliography]{revtex4-2}  

\usepackage{graphicx}
\usepackage{dcolumn}
\usepackage{bm}
\usepackage[utf8]{inputenc}
\usepackage[T1]{fontenc}
\usepackage{mathptmx}
\usepackage{afterpage}

\usepackage{color}
\usepackage{soul}

\usepackage{mathtools}

\DeclarePairedDelimiter\ket{\lvert}{\rangle}
\DeclarePairedDelimiterX\braket[2]{\langle}{\rangle}{#1 \delimsize\vert #2}

\usepackage{xcolor}
\usepackage{hyperref}
\hypersetup
{
colorlinks  = true,
linkcolor  = cyan,
citecolor = cyan,
urlcolor = magenta
}

\begin{document}

\title{Disorder-induced topological phase transition in a 1D mechanical system}

\author{Xiaotian Shi}
\affiliation{Aeronautics and Astronautics, University of Washington, Seattle, Washington 98195, USA}%

\author{Ioannis Kiorpelidis}
\affiliation{LAUM, CNRS-UMR 6613, Le Mans Université, Avenue Olivier Messiaen, 72085 Le Mans, France}%
\affiliation{Department of Physics, University of Athens, 15784, Athens, Greece}

\author{Rajesh Chaunsali}
\affiliation{LAUM, CNRS-UMR 6613, Le Mans Université, Avenue Olivier Messiaen, 72085 Le Mans, France}%

\author{Vassos Achilleos}
\affiliation{LAUM, CNRS-UMR 6613, Le Mans Université, Avenue Olivier Messiaen, 72085 Le Mans, France}%

\author{Georgios Theocharis}
\thanks{Corresponding author\\ georgiostheocharis@gmail.com}
\affiliation{LAUM, CNRS-UMR 6613, Le Mans Université, Avenue Olivier Messiaen, 72085 Le Mans, France}%

\author{Jinkyu Yang}
\thanks{Corresponding author\\ jkyang@aa.washington.edu}
\affiliation{Aeronautics and Astronautics, University of Washington, Seattle, Washington 98195, USA}%


\begin{abstract}

We numerically investigate the topological phase transition induced purely by disorder in a spring-mass chain. We employ two types of disorders -- chiral and random types -- to explore the interplay between topology and disorder. By tracking the evolution of real space topological invariants, we obtain the topological phase diagrams and demonstrate the bilateral capacity of disorder to drive topological transitions, from topologically nontrivial to trivial and vice versa. The corresponding transition is accompanied by the realization of a mechanical Topological Anderson Insulator. The findings from this study hint that the combination of disorder and topology can serve as an efficient control knob to manipulate the transfer of mechanical energy.

\end{abstract}

\maketitle

\section{Introduction}
With the discovery of Topological Insulators in condensed matter \cite{hasan2010,qi2011}, there have been tremendous efforts to explore topological phases in classical wave systems, including photonic \cite{wang2009,hafezi2013,ma2016,ozawa2019}, acoustic \cite{yang2015,he2016,ma2019}, and elastic \cite{susstrunk2015,nash2015,vila2017,xin2020} systems. Based on the bulk-edge correspondence for topological insulators, one can predict the existence of topological states on the boundary of the system by characterizing the bulk of a material with an invariant. Such states are protected by the internal symmetries and are immune to certain types and levels of disorder \cite{prodan2016}. As disorder inevitably destroys the periodicity of the system, it is natural to think that the presence of disorder tends to suppress topological properties. However, the discovery of Topological Anderson Insulators (TAI) \cite{li2009,jiang2009,groth2009,guo2010,song2014,altland2014a,mondragon-shem2014} suggests that disorder can actually induce the abnormal transition from topologically trivial to nontrivial states, which brings the studies on the interplay of topology and disorder into a new era. Later, the concept of TAI has been further illustrated and experimentally verified in atomic wires \cite{meier2018}, photonic systems \cite{stutzer2018a}, and recently in acoustic waveguides
~\cite{zangeneh-nejad2020a}. In all these cases, the tight binding model of 1D Su-Schrieffer-Heeger (SSH) was used.
However, the interplay of both topology and disorder in lattices described by a second order differential equations in time, capable of describing the dynamics of not only a variety of mechanical systems (e.g., spring-mass model) but also several electric structures (LC circuits), has not been explored.

Here, inspired by previous studies, we propose a dimer spring-mass system mimicking the 1D Su-Schrieffer-Heeger (SSH) model \cite{su1979} to realize the topological phase transition. 
Despite some similarities with previous models, this mechanical system has some important differences that can lead to distinctive behaviors in the presence of disorder. First, the band gap region is around a finite frequency as opposed to zero frequency in the SSH model. Second, in the SSH model, hopping disorder only changes the energy spectrum while preserving system's chiral symmetry. On the contrary, disorder in the spring stiffness of a spring-mass chain usually breaks the chiral symmetry, since the perturbation is reflected in the diagonal entries of the dynamical matrix as well. 
Third, as noted earlier, the equations of motions are governed by the second-order differential equations in time, such that the topological characterization of real space obtained from the transient response of the system should be handled  carefully. 
Therefore, to understand the effect of disorder on the topological properties of this classical system comprehensively, we introduce two types of disorders in our spring-mass chain: i) a chiral (chiral-symmetry preserving)  and ii) a random disorder. To probe the topological properties for various disorder scenarios, we use three kinds of topological invariants defined in the real space (displacements of the masses). Moreover, we keep track of the localization length in the system to capture the boundaries between topologically distinct phases. While the random type of disorder is more common in mechanical settings, we demonstrate that in the carefully designed setting of chiral disorder, we can achieve both kinds of topological phase transitions, from topologically nontrivial to trivial and vice versa. The later type leads to the mechanical analogue of TAI. 

\section{Model}

\subsection{Equations of Motion}

\begin{figure}[!]
	\includegraphics[width=\columnwidth]{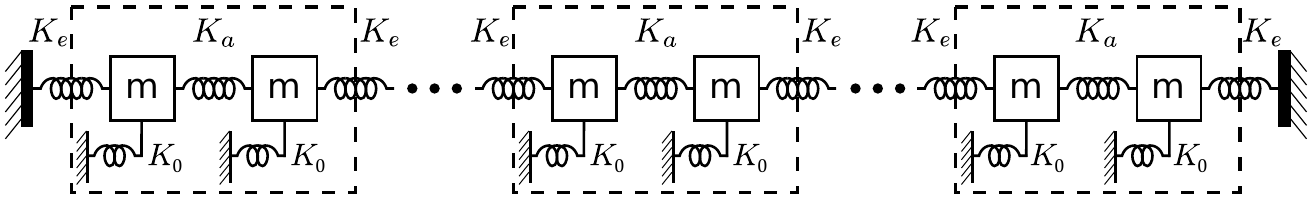}
		\caption{Schematic of the 1D dimer spring-mass chain with ground springs. Unit cell is highlighted with the box.}
	\label{FIG1}
\end{figure}

In this work, we consider a $1$D dimer spring-mass chain, as shown in Fig. \ref{FIG1}, which is composed of particles of uniform mass ($m=0.01$ throughout this study) connected by alternating springs. {All the particles are attached to the ground with an onsite spring ($K_0$), which acts in the horizontal direction.} Each unit cell contains two particles. The springs located within the unit cell are defined as intracell springs ($K_a$) while those connecting neighboring unit cells are called intercell springs ($K_e$). We assume the 1D dimer chain contains $n$ unit cells, that is $N=2n$ particles and $N+1$ inter and intracell springs in total.
By imposing fixed boundary conditions 
on both ends, the equations of motion of the system can be written as:
\begin{eqnarray}\label{eq:Eq1a}
	{m}{\ddot u_1} &=& -k_{1}u_1 - k_2(u_1-u_{2})-k_1^{0}u_1, \nonumber \\
	{m}{\ddot u_j} &=& k_{j}(u_{j - 1}-u_j) - k_{j+1}(u_j-u_{j + 1}) 
	                        - k_j^0u_j, \quad j \in [2, N-1]   \nonumber \\
	{m}{\ddot u_N} &=& k_{N}(u_{N - 1}-u_N) - k_{N+1}u_N-k_N^0u_N,
\end{eqnarray}
where $u_j$ is the displacement of the $j$th particle, $k_j$ is the spring constant between the $(j-1)$th and $j$th particles, and $k_j^0$ is the $j$th onsite spring constant, such that $k_j=K_e$ for odd $j$, $k_j=K_a$ for even $j$, and $k_j^0=K_0~\forall j$. 
%

We can rearrange the equations of motion into a matrix form as follow:
\begin{equation} \label{eq:Eq2}
\ddot{\mathbf{U}}(t)+ \mathbb{D}\mathbf{U}(t)=0,
\end{equation}
where $\mathbf{U}(t)$ and $\mathbb{D}$ are the displacement vector of length $N$ and the dynamical matrix of dimension $N \times N$, respectively. The dynamical matrix is real and symmetric and takes the following form:
\begin{widetext} 
\begin{equation}  \label{eq:Eq2a}
\mathbb{D} = \frac{1}{m}
\begin{bmatrix}
K_a+K_e+K_0 &  - K_a                        & ...                & 0 & 0 \\
- K_a                     & K_a+K_e+K_0 & - K_e          & ...  & 0 \\
... & ... & ... & ... & ... \\
0 & ... & - K_e &  K_a+K_e+K_0& -K_a\\
0 & 0 & ... & - K_a & K_a+K_e+K_0
\end{bmatrix}.
\end{equation}
\end{widetext}

To get the system's time-history response, we can rewrite Eq.~\eqref{eq:Eq2} as a first-order state equation
\begin{equation} \label{eq:Eq3}
\dot{\mathbf{X}}=\mathbb{A}\mathbf{X},
\end{equation}
\noindent where 
$\mathbf{X} 
= 
\begin{bmatrix}
\mathbf{U}(t) \\ 
\mathbf{V}(t) 
\end{bmatrix} $ 
is the state vector consisting of displacement and velocity components of all the particles, and 
$\mathbb{A} 
=  
{\begin{bmatrix}
 0&{\mathbb{I}} \\ 
 { - \mathbb{D}}&0 
\end{bmatrix}}_{2N \times 2N}$ 
with $\mathbb{I}$ being the identity matrix.

\subsection{Spectrum}

To perform the eigen analysis, we assume a harmonic system response and substitute the ansatz $\mathbf{U}(t)=\mathbf{U}e^{i \omega t}$ into Eq.~\eqref{eq:Eq2} and get:
\begin{equation} \label{eq:Eq4}
\mathbb{D}\mathbf{U}_j=\omega_j ^2\mathbf{U}_j
\end{equation}
where $\mathbf{U}_j$ is an eigenvector corresponding to the eigenfrequency $\omega_j$.
We note that after the removal of the diagonal term $(K_a+K_e+K_0)/m$ from the dynamical matrix of Eq.~\eqref{eq:Eq2a}, which shifts the spectrum to the frequency $\omega_0=\sqrt{\frac{K_a+K_e+K_0}{m}}$,
the remaining matrix obeys the anti-commutative relation:
\begin{equation} \label{eq:Eq2b}
\Gamma [\mathbb{D}-\omega_0^2 \mathbb{I}]+[\mathbb{D}-\omega_0^2 \mathbb{I}]\Gamma=0,
\end{equation}
where  $\mathbb{I}_{N \times N}$ denotes the identity matrix and 
$\Gamma$ is the chiral operator (see Appendix A for details). 
The system in this case is said to possess the chiral symmetry and in combination with the time reversal symmetry, it results in a symmetric square of the spectrum ($\omega^2$) around  the mid-gap square frequency ($\omega_0^2$).
In addition, all the eigenvectors have their chiral partner, namely for every eigenvector $\mathbf{U}_k$ with eigenfrequency  $\omega_k$, there is an eigenvector $\mathbf{U}_l$ with eigenfrequency  $\omega_l$ such that  $\Gamma \mathbf{U}_k  = \mathbf{U}_l$ and $\omega_k^2-\omega_0^2=-(\omega_l^2-\omega_0^2)$.
%

\subsection{Disorder arrangement}

Although in the clean limit (without disorder), the dynamic matrix of the dimer spring-mass chain shares the same topological property with the Hamiltonian matrix of a 1D SSH chain, these two systems become quite different in the presence of disorder. {Mathematically speaking, the diagonal terms of the SSH Hamiltonian matrix depend only on the onsite energies. 
By introducing disorder on the hoppings in the SSH tight-binding model, we will only vary the off-diagonal terms of the SSH Hamiltonian matrix.
In contrast, the diagonal terms of the dynamic matrix of a spring-mass chain are the summation of two neighboring springs and the onsite ground spring.} 
In consideration of such difference, we introduce two types of disorder to study their effects on topological phase transitions in the 1D mechanical system.
The disorder strengths of intercell and intracell springs are noted as $W_e$ and $W_a$, respectively. 
Then, the $j$th disordered spring stiffness can be written as:
\begin{equation} \label{eq:Eq5}
k_j =
    \begin{cases}
    K_e + \delta_j = K_e + W_e \epsilon_j   & \quad \text{if } j \text{ is odd}\\
    K_a + \delta_j = K_a + W_a \epsilon_j   & \quad \text{if } j \text{ is even}
  \end{cases}
\end{equation}
\noindent where $\epsilon_j$ are random, independent numbers chosen uniformly from the range $[-1,1]$. 

First, we discuss the case of chiral disorder, which we call Type I disorder in this study. For this particular type, in order to keep the chiral symmetry of the system, we choose the ground springs to take values $k_j^0=K_0+\delta_j^0$ with $\delta_j^0=-\delta_{j}-\delta_{j+1}$. By doing so, we balance out the disorders of the coupling springs, and thus the diagonal terms of the spring matrix are constant. Then, the dynamical matrix of the system with chiral disorder can be expressed as:

%
\begin{widetext} 
\begin{equation}  \label{eq:Eq6}
\mathbb{D}^{I} = \frac{1}{m}
\begin{bmatrix}
K_a+K_e+K_0 &  - {K_a} -\delta_2                        & ...                & 0 & 0 \\
- {K_a}      -\delta_2                     & K_a+K_e+K_0 & - {K_e} -\delta_3          & ...  & 0 \\
... & ... & ... & ... & ... \\
0 & ... & {- K_{e} }-\delta_{N-1} &  K_a+K_e+K_0& {- {K_a} }-\delta_N\\
0 & 0 & ... & {- {K_a}} -\delta_N & K_a+K_e+K_0
\end{bmatrix}.
\end{equation}
\end{widetext}
Note that we have again $\Gamma [{\mathbb{D}}^{I}-\omega_0^2 \mathbb{I}]+[{\mathbb{D}}^{I}-\omega_0^2 \mathbb{I}]\Gamma=0$. Therefore, the disorder matrix $\mathbb{D}^{I}-\omega_0^2 \mathbb{I}$ is chiral symmetric, as in the clean case, and a symmetric spectrum is expected to be formed around $\omega_0^2$.

We call the random disorder Type II disorder. Here, we only introduce disorders independently on intercell ($K_e$) and intracell ($K_a$) springs, while keeping the onsite springs unperturbed. In this case, the dynamical matrix takes the following form

\begin{widetext} 
\begin{equation}  \label{eq:Eq7}
\mathbb{D}^{II} = \frac{1}{m}
\begin{bmatrix}
K_a+K_e+K_0+ \Delta_1 &  - {K_a} -\delta_2                        & ...                & 0 & 0 \\
- {K_a}      -\delta_2                     & K_a+K_e+K_0 + \Delta_2 & - {K_e} -\delta_3          & ...  & 0 \\
... & ... & ... & ... & ... \\
0 & ... & {- {K_{e}}} -\delta_{N-1} &  K_a+K_e+K_0- \Delta_{N-1}&  {- {K_a} }-\delta_N\\
0 & 0 & ... & { - {K_a} }-\delta_N & K_a+K_e+K_0- \Delta_N
\end{bmatrix},
\end{equation}
\end{widetext}
where ${\Delta_{j}=\delta_{j}+\delta_{j+1}}$. As one can easily check, due to the non-constant diagonal terms, the disorder matrix $\mathbb{D}^{II}$ is not chiral, and the spectrum 
is not symmetric around the $\omega_0^2$ anymore.

\section{Topological characterization}
For an infinitely-long clean dimer chain, its topological property could be characterized by the winding number defined in the wave vector space~\cite{prodan2016}. It is quantized and can take only integer values in a system possessing chiral symmetry. However, such formula cannot be applied to the disorder system directly, since the translational symmetry is broken. Therefore, we need to handle the topological invariant in the real space. Following Ref. \cite{meier2018}, we introduce three types of topological invariants calculated by the real space wave functions, specifically \textit{Local Topological Marker} ($\nu$), \textit{mean chiral displacement} [$C(t)$] and  \textit{infinite-time limit of mean chiral displacement} ($C_{\infty}$).


\textit{Local Topological Marker (LTM)}.---This marker is based on the eigenfuctions of the system and gives a local value for the topological invariant when this is evaluated away from the boundaries~\cite{meier2018}, see the appendix of \cite{meier2018} for more information. 
First, we construct a modal matrix $\mathbb{U}$ by arranging all the normalized eigenvectors $\mathbf{U}_j$ with corresponding eigenfrequencies in ascending order. Specifically,  $\mathbb{U}=[\mathbf{U}_1, \mathbf{U}_2, ... , \mathbf{U}_n, \mathbf{U}_{n+1}, ... ,\mathbf{U}_{N}]$. Let $\mathbb{U}_{-}=[\mathbf{U}_1, \mathbf{U}_2, ... , \mathbf{U}_n]$, and $\mathbb{U}_{+}=[\mathbf{U}_{n+1}, \mathbf{U}_{n+2}, ... ,\mathbf{U}_{N}]$. Then the projectors of the negative (below the band gap) and positive (above the band gap) energy spectrum are given as $\mathbb{P}_{-}=\mathbb{U}_{-} \mathbb{U}_{-}^T$ and $\mathbb{P}_{+}=\mathbb{U}_{+} \mathbb{U}_{+}^T$, respectively.
We can then define "flat band Hamiltonian" as $\mathbb{Q} = \mathbb{P}_+ - \mathbb{P}_-$. The $\mathbb{Q}$ matrix is decomposed as $\mathbb{Q} = \mathbb{Q}_{AB}+ \mathbb{Q}_{BA}=\Gamma_A  \mathbb{Q} \Gamma_B + \Gamma_B \mathbb{Q} \Gamma_A$, where $\Gamma_A$ and $\Gamma_B$ refer to the projectors onto the $A$ or $B$ particles respectively, and $\Gamma=\Gamma_A-\Gamma_B$ is the chiral operator (see Appendix A). Then, the LTM can be defined as:

\begin{equation} \label{eq:Eq8}
\nu(l)= \frac{1}{2} \sum_{a=A,B}\{( \mathbb{Q}_{BA}[ \mathbb{X}, \mathbb{Q}_{AB}])_{la,la}+( \mathbb{Q}_{AB}[ \mathbb{Q}_{BA}, \mathbb{X}])_{la,la}\},
\end{equation}
where $\mathbb{X}$ is the position operator (see Appendix A), $l$ is the unit cell number, $lA$ and $lB$ indicate the entries of the matrix corresponding to the $A$ or $B$ particle for the $l$th unit cell. This marker works as a local topological invariant when evaluated in a region away from the boundary of the system.  To extract a value for the winding number in a disordered system, we need to average $\nu(l)$ over a small region in the center of the chain for single disorder realization ($\bar{\nu}$) then over multiple disorder realizations ($\langle \bar{\nu} \rangle$).

\textit{Mean chiral displacement (MCD)}.---{Another type of variable that can detect the winding in a 1D chiral system is the \textit{mean chiral displacement}~\cite{cardano2017,maffei2018}, given as}

\begin{equation} \label{eq:Eq9}
C(t)= 2 \langle \Psi(t) | \widehat{\Gamma} \widehat{\mathbb{X}}  |  \Psi(t) \rangle, 
\end{equation}
where $\Psi(t)=e^{-iHt}|0_a\rangle$ represents the time evolution of an initially localized state in a quantum mechanical system. Unlike the quantum mechanical system, which is governed by the Schrödinger equation, a first-order differential equation, our 1D spring-mass chain is governed by second-order differential equations. Therefore, we cannot use the above definitions of MCD directly. Instead, we first map our mechanical system to a quantum-mechanical problem. Following the framework proposed by Süsstrunk and Huber~\cite{susstrunk2016}, we introduce

\begin{equation} \label{eq:Eq10}
\Psi(t)
	= \left[ {\begin{array}{*{20}{c}}
  	{\sqrt{\mathbf{\mathbb{D}}}}&{0} \\ 
	{0} & i
\end{array}} \right] \left[ {\begin{array}{*{20}{c}}
  	{\mathbf{U}(t)} \\ 
	{\mathbf{V}(t)}
\end{array}} \right],
\end{equation}
which transform the time evolution equation given in Eq.~\eqref{eq:Eq3} into $i \dot{\Psi}(t) = \mathbb{H} \Psi(t)$ with $ \mathbb{H} = \begin{bmatrix} 0 & \sqrt{\mathbb{D} }\\ \sqrt{\mathbb{D}} & 0 \end{bmatrix}$ being a Hermitian matrix. This resembles  the Schrödinger equation. Moreover,
we define the extended chiral operator and displacement operator as $\widehat{\Gamma}=\big[\begin{smallmatrix}
  \Gamma & 0\\
  0 & \Gamma
\end{smallmatrix}\big]$ and $\widehat{\mathbb{X}}=\big[\begin{smallmatrix}
  \mathbb{X} & 0\\
  0 & \mathbb{X}
\end{smallmatrix}\big]$, respectively.  With these extended operators we can then calculate the MCD from Eq.~\eqref{eq:Eq9}.

\textit{Infinite-time-limit of mean chiral displacement (IMCD)}.---As $t \rightarrow \infty $, the MCD converges to a time-independent variable given as:

\begin{equation} \label{eq:Eq11}
C_{\infty} = 2  \sum_{j=1}^{2N} |\alpha_{aj}|^2  \langle \widehat{\Psi}_j | \widehat\Gamma  \widehat{\mathbb{X}} | \widehat{\Psi}_j \rangle
\end{equation}
 where $\widehat{\Psi}_j$ is the $j$th normalized eigenvector of the transformed Hamiltonian $\mathbb{H}$, and $\alpha_{aj}$ indicates the projection of the initial state {$\Psi(0)$} on the $j$th eigenvector.

\subsection*{Capturing topological transition through localization length}
Apart from the aforementioned topological invariants, we may also study the system's localization properties, since it is found that a topological transition is accompanied with a divergence of the localization length $\Lambda$ \cite{song2014, mondragon-shem2014} at the Fermi level, corresponding to the mid-gap frequency $\omega_0$ in our case.  
Indeed, for the chiral disorder case, at $\omega_0$ the following solutions can be obtained:
\begin{equation}
\begin{split}
u_{2n-1}=(-1)^{n-1} \prod_{j=1}^{n-1}  \dfrac{k_{2j}}{k_{2j+1}}  u_{1}  \\
u_{2n}=(-1)^{n-1} \prod_{j=1}^{n-1}  \dfrac{k_{2j+1}}{k_{2j+2}}  u_{2},
\end{split}
\end{equation}
where the $k_j$'s are given by equation (\ref{eq:Eq5}).
Then, assuming an exponential form for these solutions and applying the Birkhoff's theorem, see \cite{mondragon-shem2014}, one obtains the following expression for $\Lambda$: 

\begin{equation}
\Lambda^{-1}
=
\left|
\dfrac{1}{2} \dfrac{\int_{-1}^{1} d\epsilon \int_{-1}^{1} d\epsilon' \left( \ln|K_e+W_e \epsilon| - \ln|K_a+W_a \epsilon'| \right)}{4} 
\right|
,
\end{equation}
where an ensemble average has been used. Note that  the normalization factor $1/4$ appears since the random variables $\epsilon$ and $\epsilon'$ are uniformly distributed in the interval $[-1,1]$ as mentioned before. After performing the integration, we deduce
\begin{equation}
\Lambda^{-1}
=
\frac{1}{4} \left| 
\ln
\left[
\dfrac{\left| K_e+ W_e \right| ^ {(K_e/W_e+1)}}{\left| K_e- W_e \right| ^ {(K_e/W_e-1)}}
\dfrac{\left| K_a - W_a \right| ^ {(K_a/W_a-1)}}{\left| K_a+ W_a \right| ^ {(K_a/W_a+1)}}
\right] 
\right|
.
\label{eqLocLength}
\end{equation}
The latter expression, which is valid for the case of chiral disorder, will be used later to indicate the critical line in the ($K_a,W$) parametric plane, where  $\Lambda$ diverges.

Complementarily, we also calculate the localization length at $\omega_0$
employing the transfer matrix approach \cite{Crisanti1993,Vleck}. Specifically, we can rewrite Eq.~(\ref{eq:Eq1a})
as
\begin{equation}
\begin{pmatrix}
u_{j+2} \\ u_{j+1}
\end{pmatrix}
=
\mathbb{T}_j
\begin{pmatrix}
u_{j+1} \\ u_{j}
\end{pmatrix},
\end{equation}
and get the transfer matrix $\mathbb{T}_j$ given by
\begin{equation}
\mathbb{T}_j
=
\begin{pmatrix}
\dfrac{k_{j+2} + k_{j+1} + k_{j+1}^0 -m\omega_0^2 }{k_{j+2}} && -\dfrac{k_{j+1}}{k_{j+2}} \\
1 && 0
\end{pmatrix}
.
\label{eq12}
\end{equation}
Then, the Lyapunov exponents $\gamma_1$ and $\gamma_2$ are calculated numerically using the typical numerical schemes described in \cite{MacKinnon1981,MacKinnon1983,Vleck}. We found that $\gamma_1\approx -\gamma_2=\gamma$.
The localization length is then determined through the relation
$\Lambda=\dfrac{1}{\gamma}$.

\section{Numerical Results}

\subsection{No disorder}
\begin{figure}[!]
\includegraphics[width=\columnwidth]{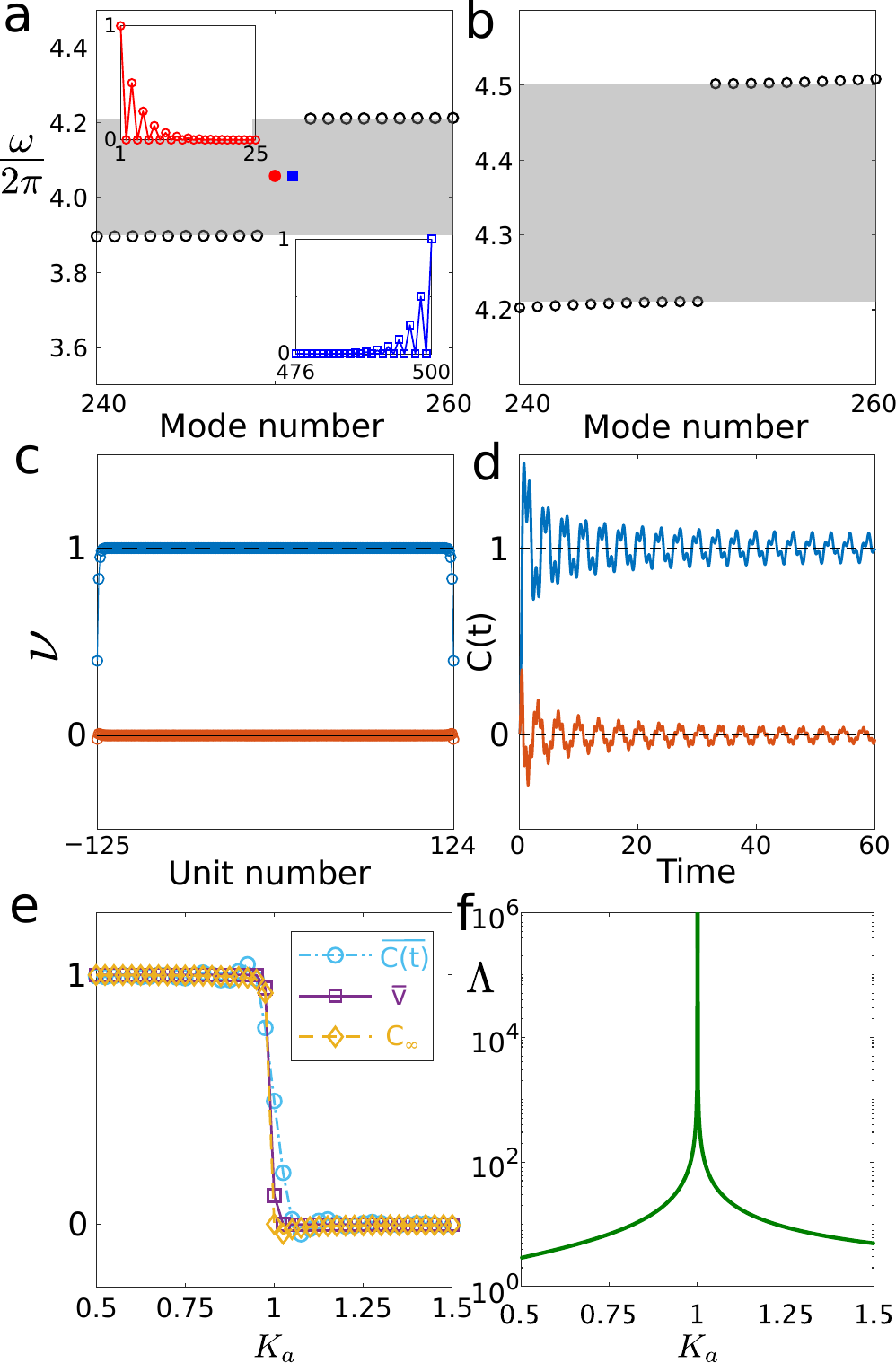}
\caption{Eigenfrequencies for a finite (a) topologically nontrivial and (b) trivial 1D spring-mass chain. Shapes of the topological edge modes are shown in the inset of (a). The grey areas represent the band gaps. (c) LTM calculated from the eigenmodes of the nontrivial (blue) and trivial (red) system. (d) Time-dependent MCD for the nontrivial (blue) and trivial (red) under initial displacement excitation. (e) Evolution of three kind of real space topological invariants during a topological transition. (f) Localization length of the eigenmode at the frequency $\omega_{0}$. 
}
\label{FIG2}
\end{figure}
Before moving to the disorder study, we need to verify the aforementioned three types of real space topological invariants in a finite clean system. We consider a chain composed of $n=250$ unit cells, i.e., $500$ particles and $501$ inter/intracell springs. In the following numerical simulations, we fix  $K_0=5$. First, let us look at a case with $K_e=1$ and $K_a=0.5$. For $K_e>K_a$, the periodic system is topological, which is characterized by a winding number $v=1$. In a finite chain, we expect to see two localized edge modes exist in the topological band gap as shown in Fig.~{\ref{FIG2}}(a)
[see also Fig.~{\ref{FIG7}}(a) in Appendix B for the whole spectrum].
The red and blue dots represent the two states localized on the left,  $\ket{L}$, and right boundary, $\ket{R}$, respectively
at the mid gap frequency $\omega_0$. In fact, a hybrization of the edge states is expected, which leads to a frequency splitting, meaning that the two eigenfrequencies are not exactly located at $\omega_0$ but at $\omega_{0}^{+}$ and $\omega_{0}^{-}$ (the two corresponding eigenstates are approximated as:
$\ket{\omega_{0}^{+}}=\frac{e^{-i\phi/2}\ket{L}+e^{i\phi/2}\ket{R}}{\sqrt{2}}$ and $\ket{\omega_{0}^{-}}=\frac{e^{-i\phi/2}\ket{L}-e^{i\phi/2}\ket{R}}{\sqrt{2}}$ for some $\phi$) \cite{Asboth2016}. However, for large system sizes, as in our case, one can consider that the energy splitting is sufficiently small, and thus we have a "degeneracy" of the eigenstates  $\ket{L}$ and  $\ket{R}$ at $\omega_{0}$.
The shapes of the localized edge modes are shown in the insets of Fig.~{\ref{FIG2}}(a). Note that the two localized edge modes have a characteristic profile with only sites from one sub-lattice excited (either $j$ even or odd). 
This is a well-known consequence of the chiral symmetry.
As we increase $K_a$ to $1.5$, the system becomes trivial with a winding number being zero, where the localized edge modes do not exist anymore [see Fig.~{\ref{FIG2}}(b) and Fig.~{\ref{FIG7}}(b) in Appendix B].

Based on the eigenmodes obtained from Eq.~\eqref{eq:Eq4}, we can calculate the LTM of the system as defined in Eq.~\eqref{eq:Eq8}. Figure~{\ref{FIG2}}(c) shows the calculated LTM as a function of the unit cell number in the topologically nontrivial (red line) and trivial cases (blue line), respectively. As we see, the LTM is well quantized to either zero or one in the bulk of the chain, while approaching the boundary, it deviates from the theoretical value. To overcome such boundary effect, we take the average value over 100 unit cells in the middle of the chain, noted as $\bar{v}$. We then get $\bar{v}=1$ for the nontrivial and $\bar{v}=0$ for the trivial configurations. Similarly, by plugging the eigenmodes into Eq.~\eqref{eq:Eq11}, we obtain the IMCD, which are $C_{\inf}=1$ and $C_{\inf}=0$ for nontrivial and trivial configurations, respectively. To derive the time-dependent MCD, we take the initial displacement $u_{251}(0)=0.1$ and perform transient simulations. After normalizing the time-history response at each time $t$ and plugging this in Eq.~\eqref{eq:Eq9}, we get the MCD as a function of time. Figure~{\ref{FIG2}}(d) demonstrates its evolution in a nontrivial (blue curve) or trivial (red curve) system. {It is clear that the MCD oscillates and tends to converge to 1 (0) in a topologically nontrivial (trivial) system despite the oscillations. We conduct the time average from $t=0$ to $t=60$ to eliminate the influence of fluctuations, which gives $\bar{C}(t)=1$ [$\bar{C}(t)=0$]. The time window is selected to avoid possible wave reflections at the system boundaries. 

In these two cases, 
all three topological invariants are closely quantized to $0$ or $1$, which shows a great agreement with the theoretical winding number of an infinite system defined in the momentum space. Figure~{\ref{FIG2}}(e) summarizes the evolution of the different topological invariants during an \textit{in} \textit{situ} topological phase transition. During the process, we vary $K_a$ from $0.5$ to $1.5$ with a $0.025$ step while keeping $K_e=1$. 
Clearly, all three kinds of real space topological invariants agree well with each other and can accurately probe the topological phase transition in a clean 1D mechanical system, which happens when the gap closes, namely at $K_a=K_e=1$. 
Finally, in Fig.~\ref{FIG2}(f), we also plot the the localization length of the edge modes as we vary the coupling $K_a$ from 0.5 to 1.5,  which is given by the analytical expression $\Lambda= \frac{2}{|\ln(K_a/K_e)|}$.
Notice that at the point where the topological transition occurs, i.e., $K_a=K_e=1$, the localization length diverges. This is expected since, as we mentioned, the topological transition occurs when the gap closes and thus the edge mode transforms to an extended mode.

\subsection{Chiral disorder}
{In this subsection, we explore the topological phase transition of a spring-mass system induced by the chiral disorder with the dynamical matrix $\mathbb{D}^I$. 
Figure~{\ref{FIG3}}(a) shows the numerically calculated LTM as a function of the intracell spring stiffness $K_a$ and the disorder strength $W=W_a=2W_e$ under chiral disorder. We again choose $K_e=1$. Each data point is a disorder-averaged result over $250$ realizations. Evidently, the surface map can be approximately divided into two regions separated by a blurry white boundary. The region with a red (blue) tone has topological marker close to $1$ ($0$) corresponding to a topologically nontrivial (trivial) system. In most real mechanical systems, the spring stiffness tend to be positive. The black dashed lines follow the relation given by $\mathrm{min}(|K_e-W_e|, |K_a-W_a|)=0$ for sufficiently large $K_0$. Therefore, only the area on the left of the dashed black line in the surface plot could be realized without the necessity of negative-stiffness springs. However, note that by tuning the disorder ratio ($W_e/W_a$), we can further shift the topological phase diagram and change the critical boundary (see Appendix C). Note also that effective negative-stiffness springs can be designed utilizing the symmetry of modes and a sophisticated geometry of the coupling of the neighbor particles, look for example refs.~\cite{Serra-Garcia2018} and \cite{Chen2020} for an implementation of negative hoppings in elastic and acoustic systems, respectively.

\begin{figure}[!]
\includegraphics[width=\columnwidth]{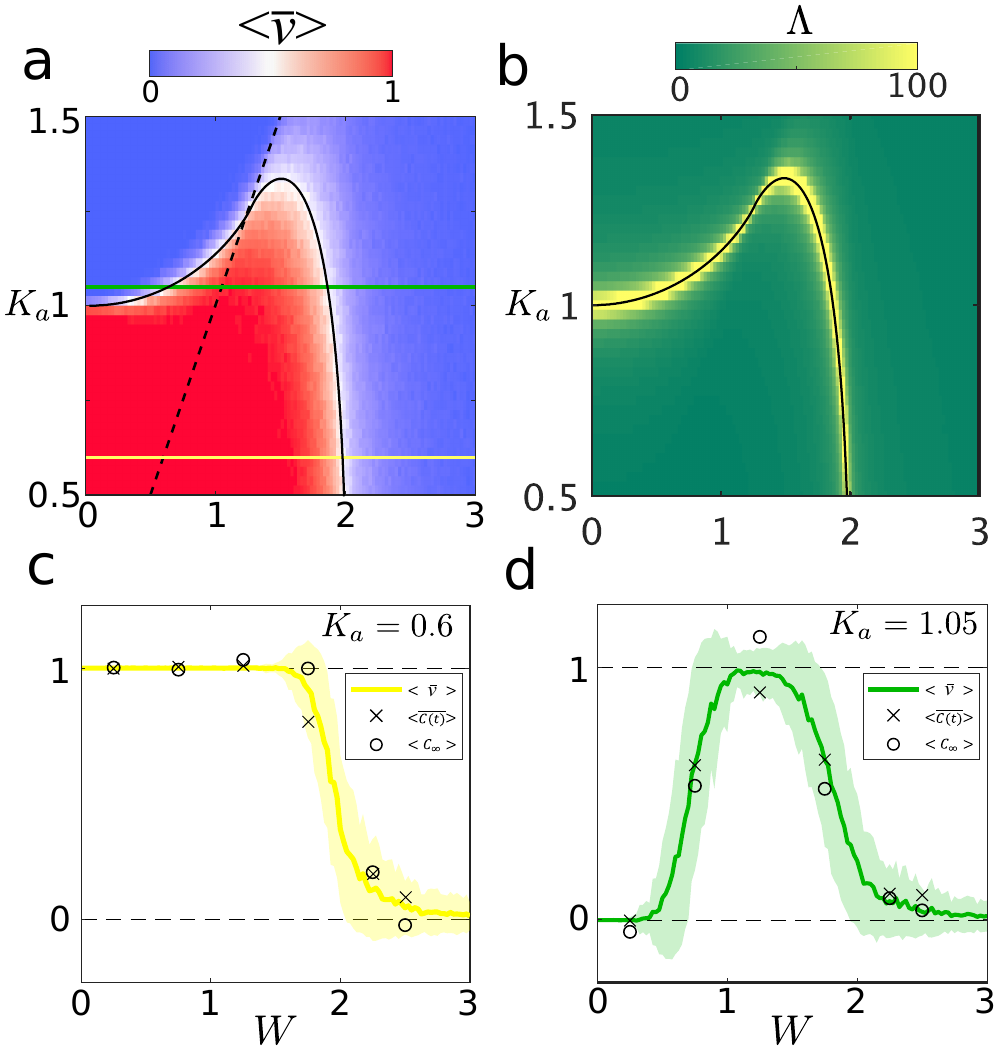}
\caption{(a) Topological phase diagram under chiral disorder with disorder ratio $W=W_a=2W_e$. Colorbar ranging from $0$ to $1$ stands for the LTM averaged over $250$ disorder realizations. 
(b) The localization length at the mid gap frequency, $\Lambda(\omega_0)$ from the transfer matrix method (colormap) and the analytical expression.
(c) and (d) The evolution of topological markers with the increase in disorder strength for $K_a=0.6$ [yellow line in (a)]  and $1.05$ [green line in (a)], respectively.  The lines and shaded area represent the mean and standard deviation of the LTM. The cross and sphere markers indicate the mean of MCD and IMCD.}
\label{FIG3}
\end{figure}

Figure~\ref{FIG3}(b) shows the localization length at $\omega_{0}$, obtained numerically using the transfer matrix method, as a function of the same parameters ($K_a$ and $W$). We note here that the transfer matrix was iterated $10^6$ times at each point of Fig.~{\ref{FIG3}}(b). The solid black curve shown in Figs.~{\ref{FIG3}}(a) and {\ref{FIG3}}(b) indicates the critical boundary obtained by the analytical diverging line from Eq.~({\ref{eqLocLength}}). It is clear that the numerical solutions [blurry white regions in Fig.~{\ref{FIG3}}(a) and highlighted area in Fig.~{\ref{FIG3}}(b)] match with the analytical results (solid black curve). This denotes that the topological phase transition is related with a critical point at which the localization length diverges.

Figure~\ref{FIG3}(c) shows the evolution of the topological marker with the increase of the disorder strength for $K_a=0.6$, which corresponds to the yellow line in Fig.~\ref{FIG3}(a). In the clean limit ($W=0$), this finite system is nontrivial with a topological marker of $\bar{v}=1$. As expected, the topological marker remains constant in the presence of weak and medium level disorder, as the topological properties of this system are protected by the chiral symmetry and the topological edge modes are immune to weak disorder. However, when the disorder strength rises and reaches a sufficiently large amount, the topological marker falls sharply and eventually stabilizes near zero under strong disorder, which indicates a transition process from a topologically nontrivial system to a trivial one. We also notice that the disorder-averaged results show larger standard deviation in this region, which is reasonable in strong disorder. We also show in the same figures disorder-averaged mean values of MCD and IMCD, marked by the cross and sphere markers. See Appendix D for typical time-history responses of the 1D spring mass chain under different levels of disorder. 
All three kinds of topological invariants therefore show good agreement in capturing the general trends of disorder-induced topological transition. 
For the time-history response, note that negative stiffness of intersite or onsite springs caused by the strong disorder does not necessarily lead to dynamical instability of our system. Within the disorder ranges that we are interested in, our system remains stable (See Appendix E for more details).


For $K_a=1.05$ [green line in Fig.~{\ref{FIG3}}(a)], however, we observe a more sophisticated topological phase transition process, as shown in Fig.~{\ref{FIG3}}(d). The system is trivial with $\bar{v}=0$ when there is no disorder. For very weak disorder $W<0.3$, the topological marker remains near zero. However, as we introduce stronger disorder, the topological marker surprisingly surges up to reach a plateau ($\bar{v} \approx 0.98$) which is very close to $1$ (not exactly 1 due to finite size). If we keep increasing the disorder strength, it drops again close to zero. This procedure is particularity interesting since we demonstrate a two-way transition by purely increasing disorder in the system. The spring-mass chain starts as a trivial system, then changes into a nontrivial one and finally becomes a trivial Anderson insulator. The MCD and IMCD results as marked with crosses and circles in Fig.~{\ref{FIG3}}(d) also corroborate the LTM trend. Therefore, the intermediary nontrivial state is the realization of a mechanical analogue to the TAI.

\begin{figure}[!]
\includegraphics[width=\columnwidth]{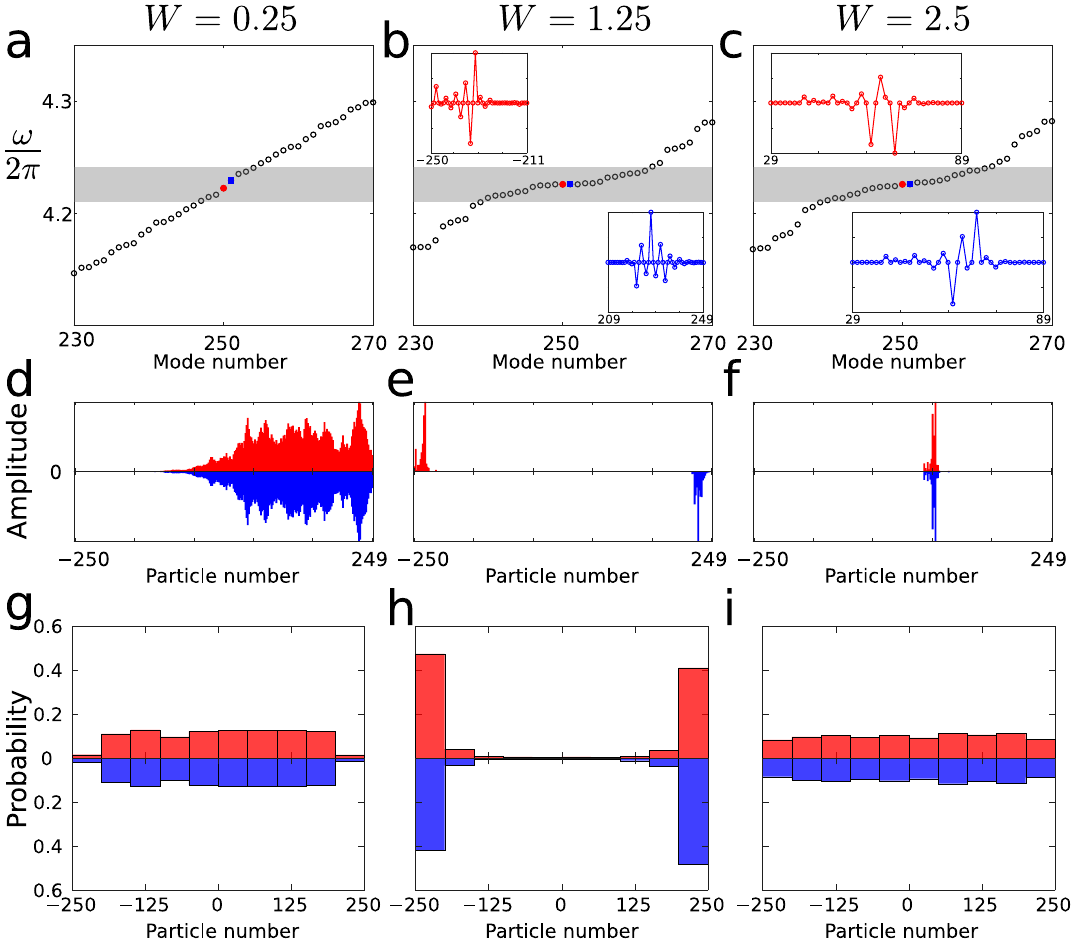}
\caption{Eigen-frequencies and two modes in the center of frequency spectrum of a finite spring-mass chain ($K_a=1.05$ and $K_e=1$) under \textit{chiral} disorder with disorder ratio $W=W_a=2W_e$ for (a) $W=0.25$, (b) $W=1.25$ (TAI phase), and (c) $W=2.5$. These correspond to three typical realizations with different disorder strengths in Fig.~\ref{FIG3}(d). Chiral nature of the two edge modes lying at the center of band gap is shown in the insets.
(d)--(f) Spatial profile of these two modes for one disorder realization.
(g)-(h) Histogram of the Center of Mode (COM) obtained by $1000$ disorder realizations, describing the probability of COM lying in 50 particle interval throughout the chain.}
\label{FIG4}
\end{figure}

So far, we have quantitatively demonstrated the topological phase transitions by tracking the real-space topological invariants. Now we investigate their characteristics in more details. We choose the configuration of the system that supports the TAI phase, i.e., with $K_a=1.05$ and $K_e=1$ in Fig.~{\ref{FIG3}}(d) and analyze its eigenfrequencies and modes as a function of disorder strength. 
Figures~{\ref{FIG4}}(a)-(c) show the eigenfrequencies of a finite chain under one realization with disorder strengths $W=0.25$, $W=1.25$, and $W=2.5$. The grey regions represent the topological band gaps in the clean ($W=0$) case. After we introduce disorder in the system, the topological band gap shrinks in the regime of weak disorder [Fig.~{\ref{FIG4}}(a) for $W=0.25$] and is eventually closed under strong disorder [see Figs.~{\ref{FIG4}}(b) and {\ref{FIG4}}(c) for $W=1.25$ and $W=2.5$, respectively]. 
We then plot the shapes of the center modes ($250$th and $251$st) in Figs.~{\ref{FIG4}}(d)-(f). We find that for weak disorder ($W=0.25$), the modes are quite extended. Of course, as the disorder strength increases, the modes become spatially strongly localized. 
The TAI phase ($W=1.25$) is distinct in a way that the localized modes appear on the boundaries of the system as in a nontrivial system, whereas the localized modes need not be at the boundaries for an Anderson insulator ($W=2.5$) and reside at the same location with only phase differences at some particles. 

We further demonstrate the validity of this statement for all disorder realizations by defining the Center of Mode (COM) as
\begin{equation} \label{eq:Eq12}
\mathrm{COM} = \frac{\displaystyle\sum_{i=1}^{N} x_i {{u}_i^2}}{\displaystyle\sum_{i=1}^{N} { u_i^2}}
\end{equation}
where $x_i$ and $u_i$ denote the position and displacement of the $i$th mass. We study the statistical distribution of the COM for $1000$ disorder realizations.
Figures ~{\ref{FIG4}} (g)-(i) show the histograms of the COM of the 250th and 251st modes under chiral disorder for varying disorder strengths.  
This agrees with the topological phase diagram [see Fig.~\ref{FIG3}(d)] and the eigen analysis [Figs.~{\ref{FIG4}} (d)-(f)]. When the system is with weak ($W=0.25$) and strong ($W=2.5$) disorder, the 250th and 251st modes are likely to be all along the chain; however, for the TAI phase ($W=1.25$), the modes are most likely localized at the boundaries [Fig.~{\ref{FIG4}} (h)]. A similar inference can be made for the configuration with  $K_a=0.6$ and $K_e=1$ and that has been included in Appendix F.

\subsection{Random disorder}
\begin{figure}[!]
\includegraphics[width=\columnwidth]{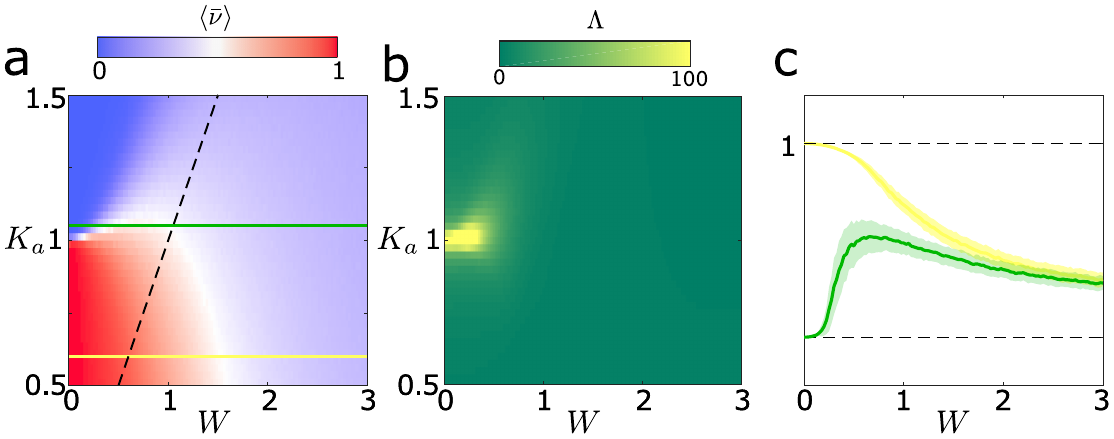}
\caption{(a) Topological phase diagram under random disorder with disorder ratio $W=W_a=2W_e$. Colorbar ranging from $0$ to $1$ stands for the LTM  averaged over $250$ disorder realizations. (b) Localization length $\Lambda(\omega_{\text{0}})$ under random disorder. The transfer matrix was again iterated $10^6$ times at each point. (c) The evolution of topological marker with increase of disorder for $K_a=1.05$ (green) and $K_a=0.6$ (yellow). The shaded area shows the standard deviation.}
\label{FIG5}
\end{figure}
We now discuss the effect of random disorder on topology by removing the disorder on the onsite ground springs. As given in Eq.~\eqref{eq:Eq7}, the diagonal entries of the random disordered stiffness matrix are not identical anymore, which leads to the breaking of the system's chiral symmetry. Since the topology in the 1D SSH model is mainly protected by the chiral symmetry, we expect to see a stronger influence of the random disorder over topology.
Figure~{\ref{FIG5}}(a) shows the topological phase diagram under random disorder with $W=W_a=2W_e$. It is clear that the topological states can only survive much weaker random disorder compared to the chiral one. Moreover, Fig.~\ref{FIG5}(b) shows the localization length at $\omega_{\text{0}}$, and in contrast to the case of chiral disorder there is no signature of the localization length divergence at this frequency, which also contributes to the uncertainty of the clear phase transition due to non-chiral disorder.
Furthermore, even with very small amount of disorder, we see that the topological marker deviates from the quantized value as shown in Fig.~{\ref{FIG5}}(c). As we expected, breaking the chiral symmetry leads to the variation of the real space topological marker, which is not well quantized to an integer value but tends to saturate near a decimal between $0$ and $1$. This leave us an open question whether it is appropriate to discuss the concept of topology in a 1D mechanical system in absent of the chiral symmetry. If so, what kind of quantities or phenomena should we use to characterize the topological properties in these cases? 

\begin{figure}[!]
\includegraphics[width=\columnwidth]{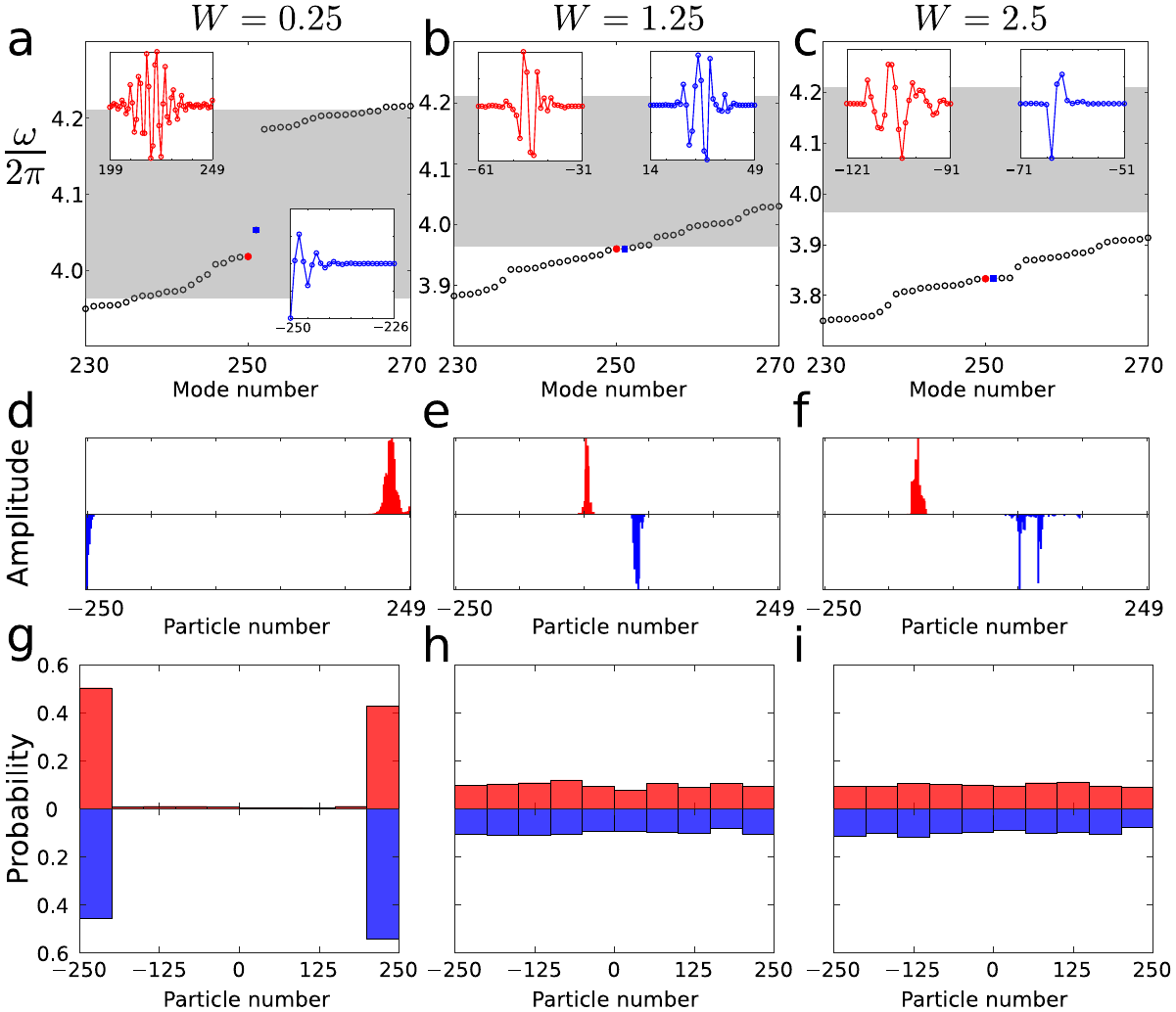}
\caption{Eigen-frequencies and two modes in the center of frequency spectrum of a finite spring-mass chain ($K_a=0.6$ and $K_e=1$) under \textit{random} disorder with disorder ratio $W=W_a=2W_e$ for (a) $W=0.25$, (b) $W=1.25$, and (c) $W=2.5$. These correspond to three typical realizations with different disorder strengths on the yellow curve in Fig.~\ref{FIG5}(c). Non-chiral nature of the two modes (marked) is shown in the insets.
(d)--(f) Spatial profile of these two modes for one disorder realization.
(g)-(h) Histogram of the Center of Mode (COM) obtained by $1000$ disorder realizations, describing the probability of COM lying in 50 particle interval throughout the chain.}
\label{FIG6}
\end{figure}

Despite the lack of a theoretical framework, we try to extend the eigen analysis and COM statistical study to examine the relation between random disorder and the topological indicators in a 1D system. Figures~{\ref{FIG6}}(a)-(c) show the eigenfrequencies around the center of the frequency spectrum of a finite chain ($K_a=0.6$) under random disorder with strength $W=0.25$, $W=1.25$, and $W=2.5$, respectively. The grey regions represent the topologically nontrivial band gaps in the clean ($W=0$) case. It is clear that the random disorder causes more drastic change of the frequency spectrum. Especially, the two center modes no longer share a common frequency and may shift out of the original bandgap region. Despite the frequency variations, the two center modes of the system under weak ($W=0.25$) random disorder are still localized modes existing at the boundary [Figs.~{\ref{FIG6}}(d)-(i)], which is the key character of the topological edge states. This is consistent with the topological invariant calculation, which is very close to $1$ in this configuration [see the yellow line in Fig.~{\ref{FIG5}}(c)]. As we increase the disorder level, these center modes start to move into the bulk. More interestingly, the two center modes tend to localize at different positions in the chain, which is different from the trivial chiral disordered system where the two modes exist at the same location [Fig.~\ref{FIG4}(f)]. A similar trend can also be observed for the disordered study in the case of $K_a=1.05$ (Appendix F). 
Finally, in Figs.~{\ref{FIG4}} (g)-(i) we show the histograms of the COM of the central modes under random disorder for varying disorder strengths. For weak ($W=0.25$) disorder, the 250th and 251st modes are likely to be localized at the boundaries, while for strong ($W=1.25$ and $W=2.5$) disorder (both in the trivial Anderson phase) they can be localized all along the chain.

\section{Conclusion}
In this work, we explore the disorder-induced topological phase transitions in a spring-mass chain. To probe the topological property in a disordered 1D mechanical system, we use three kinds of topological invariants obtained from the real space wave functions to work as an effective winding number. We introduce two types of disorder, chiral and random, to study their influences on the topology in a mechanical system. Under the chiral disorder, we not only demonstrate the transition from topological to trivial, but also the abnormal reverse transformation leading to the realization of so-called Topological Anderson Insulator (TAI) in a mechanical setup. However, for the case of random disorder, we do not observe similar TAI phenomena. Instead, we find that the random disorder suppresses the topological property, as it breaks the internal system symmetry that protects the topology. Our findings can be extended to higher dimensional mechanical systems to further study the interactions between disorder and topology. In addition, despite the difficulties in controlling stiffness in a disordered mechanical system, it would be very interesting to implement and experimentally verify the proposed framework in other highly tunable systems, such as electric circuits~\cite{zhang2019a}.

\begin{acknowledgments}
We thank Professor Christopher Chong from Bowdoin College for fruitful discussions. X. S., and J. Y. are grateful for the financial support from the U.S. National Science Foundation (CAREER1553202 and EFRI-1741685). I.K. acknowledges financial support from the Academy of Athens. G.T. acknowledges financial support from the CS.MICRO project funded under the program Etoiles Montantes of the Region Pays de la Loire.
\end{acknowledgments}

\section*{APPENDIX A: Definition of the chiral operator and position operator}

{For a 1D system with $n=250$ unit cells, that is $N=2n=500$ particles, we first define the unit cell numbers as $l=[-125,-124,\cdots, 0, \cdots,123, 124]$.} Then the chiral operator and position operator are given as

\begin{align}
\Gamma = {\left[ {\begin{array}{*{20}{c}}
  {1}& {0}& 0 & 0 & \cdots\\ 
  {0}& {-1} & 0 & 0 & \cdots\\ 
  {0}& {0}& {1} & 0 & \cdots\\ 
  {0}& {0}& {0} & {-1} & \cdots\\ 
  \vdots & \vdots & \vdots & \vdots & \ddots
\end{array}} \right]_{N \times N}},
\end{align}

\begin{align}
\Gamma_A = {\left[ {\begin{array}{*{20}{c}}
  {1}& {0}& 0 & 0 & \cdots\\ 
  {0}& {0} & 0 & 0 & \cdots\\ 
  {0}& {0}& {1} & 0 & \cdots\\ 
  {0}& {0}& {0} & {0} & \cdots\\ 
  \vdots & \vdots & \vdots & \vdots & \ddots
\end{array}} \right]_{N \times N}}, &
\Gamma_B = {\left[ {\begin{array}{*{20}{c}}
  {0}& {0}& 0 & 0 & \cdots\\ 
  {0}& {1} & 0 & 0 & \cdots\\ 
  {0}& {0}& {0} & 0 & \cdots\\ 
  {0}& {0}& {0} & {1} & \cdots\\ 
  \vdots & \vdots & \vdots & \vdots & \ddots
\end{array}} \right]_{N \times N}},
\end{align}

\begin{align}
\mathbb{X} = {\left[ {\begin{array}{*{20}{c}}
  {-125}& {0}& 0 & 0 & \cdots & 0 & 0\\ 
  {0}& {-125}& 0 & 0 & \cdots & 0 & 0\\ 
  {0}& {0}& {-124} & 0 & \cdots & 0 & 0\\ 
  {0}& {0}& {0} & {-124} & \cdots & 0 & 0\\ 
  \vdots & \vdots & \vdots & \vdots & \ddots & \vdots & \vdots\\
  {0}& {0}& 0 & 0 & \cdots & 124 & 0 \\
  {0}& {0}& 0 & 0 & \cdots &  0& 124 \\
\end{array}} \right]_{N \times N}}.
\end{align}

\section*{APPENDIX B: Spectrum}
	\begin{figure}[!]
	\includegraphics[width=\columnwidth]{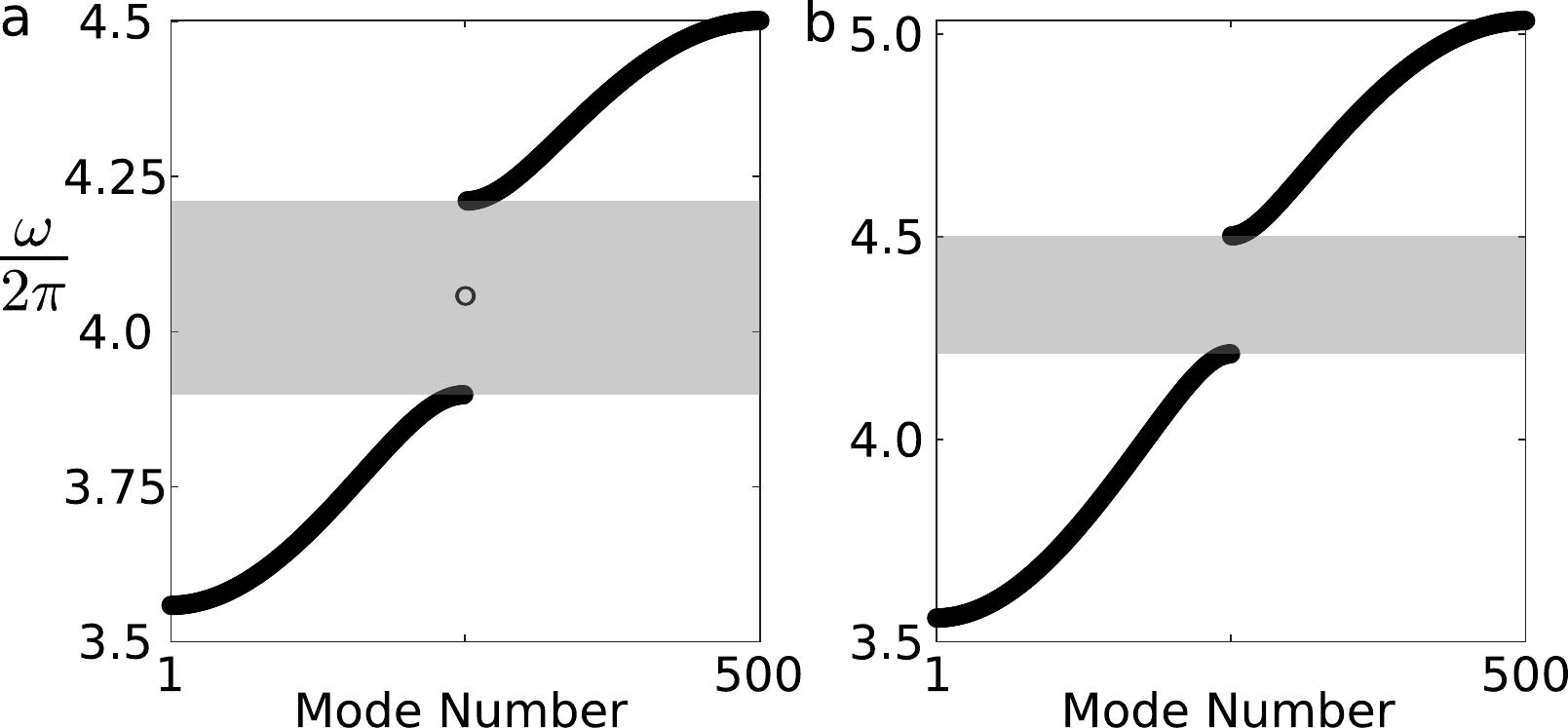}
		\caption{The eigenfrequencies of a chain of $500$ particles for the topological nontrivial phase (left panel, $K_e=1$, $K_a=0.5$) and trivial case (right panel,  $K_e=1$, $K_a=1.5$). For both cases, $K_0= 5$. Band gap is highlighted with the shaded area.}
	\label{FIG7}
\end{figure}

In Fig.~\ref{FIG7}, we show \textit{all} the eigenfrequencies of a clean chain of $500$ particles for both topologically trivial and nontrivial phases. These are the same cases as shown in Fig.~\ref{FIG2}a and \ref{FIG2}b of the main text.

\section*{APPENDIX C: Topological phase diagram with different combinations of disorder strength }
{In Section IV-B of the main text, we discuss the disorder configuration with specific ratio $W=W_a=2W_e$. Here, we briefly explore the effect of disorder ratio ($W_e/W_a$) by studying the topological transitions in the presence of chiral disorder with different combination of $W_a$ and $W_e$. Figures~{\ref{FIG8}}(a)-(c) show the topological phase diagrams under chiral disorder with (a) $W=W_a$ and $W_e=0$, (b) $W=W_a=4W_e$, and (c) $W=W_a=0.5W_e$. It is clear that by changing the disorder ratio ($W_e/W_a$), we can shift the critical boundaries for the disorder-induced topological transitions.
\begin{figure}[h]
\includegraphics[width=\columnwidth]{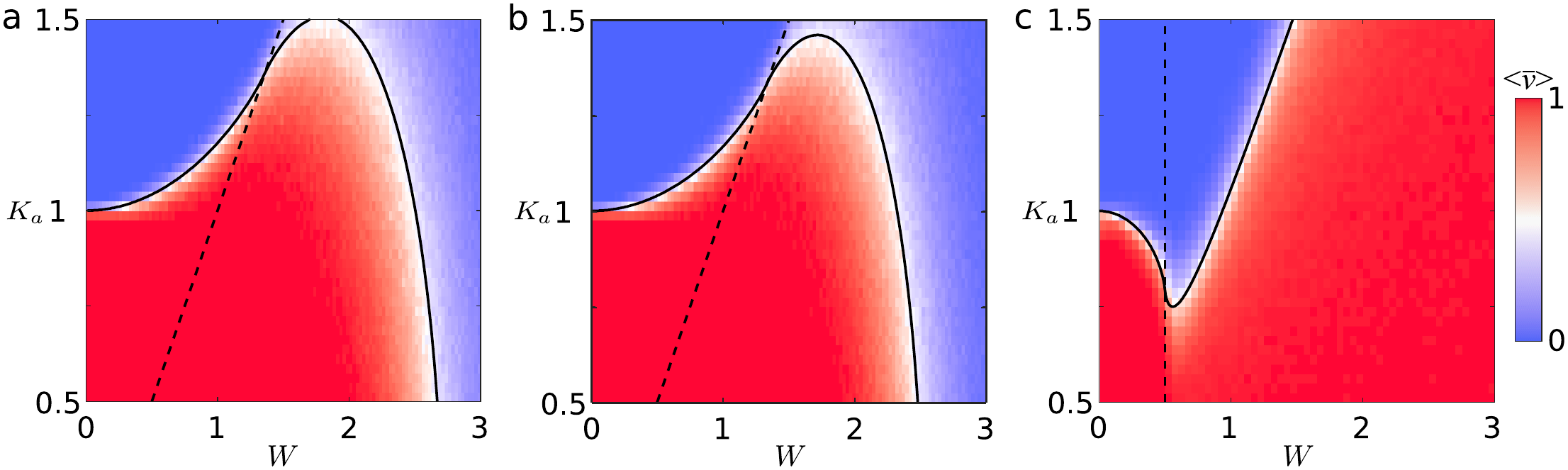}
\caption{Topological phase diagram under chiral disorder with different combination of $W_a$ and $W_e$. (a) $W=W_a$ and $W_e=0$, (b) $W=W_a=4W_e$, and (c) $W=W_a=0.5W_e$. The solid black line indicates the divergence of the localization length obtained analytically. The dashed line separates the region of negative stiffness.}
\label{FIG8}
\end{figure}

\section*{APPENDIX D: Spatiotemporal diagrams in disordered systems}
\begin{figure}[h]
\includegraphics[width=\columnwidth]{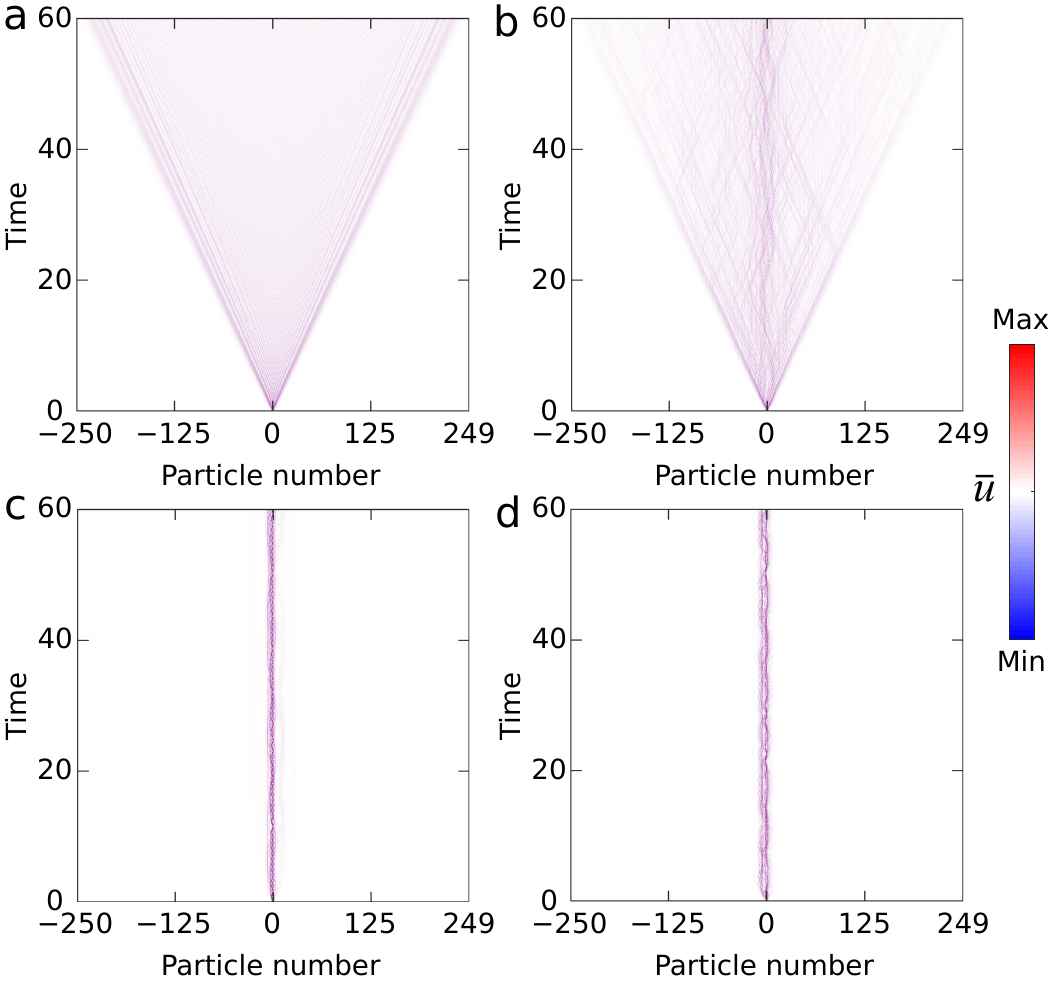}
\caption{Spatio-temporal evolution of particle displacement ($K_a=0.6$ and $K_e=1$). We use chiral-disorder strengths (a) $W=0$, (b) $W=0.25$, (c) $W=1.25$, and (d) $W=2.5$.}
\label{FIG9}
\end{figure}
Figures~{\ref{FIG9}}(a)-(d) show the typical time-history responses of the 1D spring mass chain with chiral disorder under strength $W=0$, $W=0.25$, $W=1.25$, and $W=2.5$, respectively. In a periodic system ($W=0$), the energy is evenly spreading to both sides of the system after the initial disturbance in the middle of the chain, as shown in Fig.~{\ref{FIG9}}(a). With the increase of disorder strength, we start to see more and more localization of the energy in the system due to Anderson localization.

\begin{figure}[!]
\includegraphics[width=0.6\columnwidth]{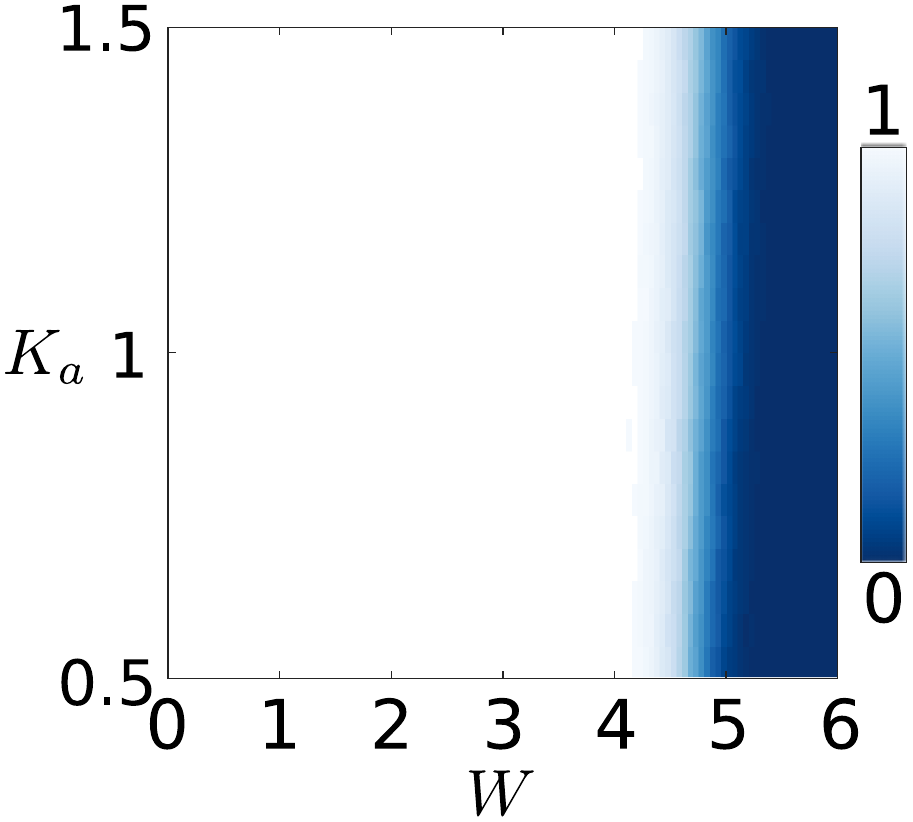}
\caption{Stability diagram of the 1D system as a function of $K_a$ and $W$ under the chiral disorder. Colorbar ranging from $0$ to $1$ stands for the probability (over $1000$ disorder realizations) of the dynamical matrix being positive-definite.}
\label{FIG10}
\end{figure}

\section*{APPENDIX E: Stability of disordered systems}
The dynamics of our system is governed by Eq.~{\eqref{eq:Eq2}}. For the stability of this system, the dynamical matrix must be positive definite, implying all the eigenvalues ($\omega^2$) are positive. In Fig.~{\ref{FIG10}}, we show the stability diagram under chiral disorder that correspond to the case shown in Fig.~\ref{FIG3}(a) of the main text. This highlights the probability of the dynamical matrix being positive-definite with varying system parameters. We can therefore conclude that our system remains stable within the disorder range we are interested in ($W<3$).

\section*{APPENDIX F: Some other cases of chiral and random disorder}
To have a comprehensive understanding of the topological transition process, we study two extra cases with chiral and random disorders, respectively. In Fig.~{\ref{FIG11}}, we study a spring-mass chain with configuration $K_a=0.6$ and $K_e=1$ under \textit{chiral} disorder, which corresponds to Fig.~{\ref{FIG3}} (c). Similarly, Fig.~{\ref{FIG12}} shows the results of a system with configuration $K_a=1.05$ and $K_e=1$ under \textit{random} disorder corresponding to the green curve in  Fig.~{\ref{FIG5} (c)}.

\begin{figure}[!]
\includegraphics[width=\columnwidth]{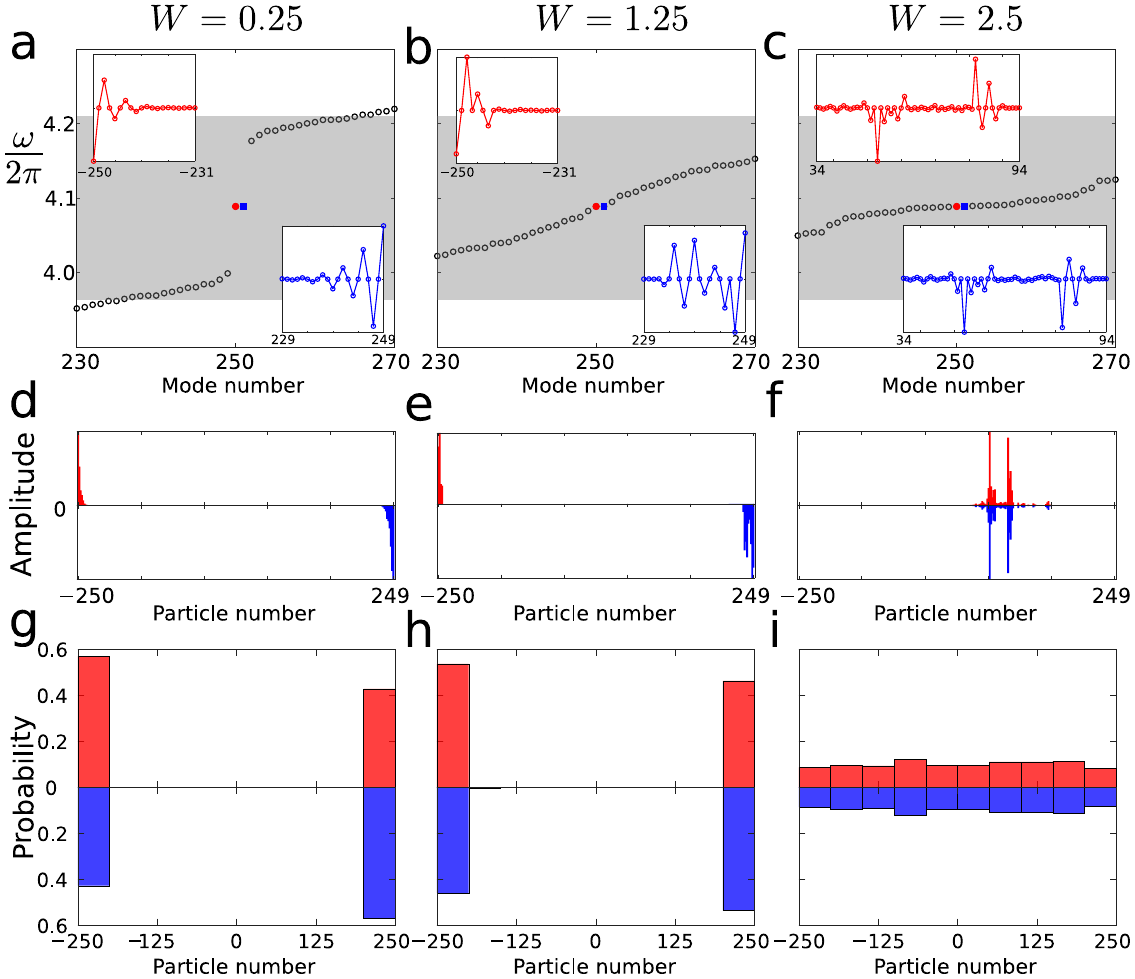}
\caption{Eigen-frequencies and two modes in the center of frequency spectrum of a finite spring-mass chain ($K_a=0.6$ and $K_e=1$) under \textit{chiral} disorder with disorder ratio $W=W_a=2W_e$ for (a) $W=0.25$, (b) $W=1.25$, and (c) $W=2.5$. These correspond to three typical realizations with different disorder strengths in Fig.~\ref{FIG3}(c). Chiral nature of the two edge modes lying at the center of band gap is shown in the insets.
(d)--(f) Spatial profile of these two modes for one disorder realization.
(g)-(h) Histogram of the Center of Mode (COM) obtained by $1000$ disorder realizations, describing the probability of COM lying in 50 particle interval throughout the chain.}
\label{FIG11}
\end{figure}

\begin{figure}[!]
\includegraphics[width=\columnwidth]{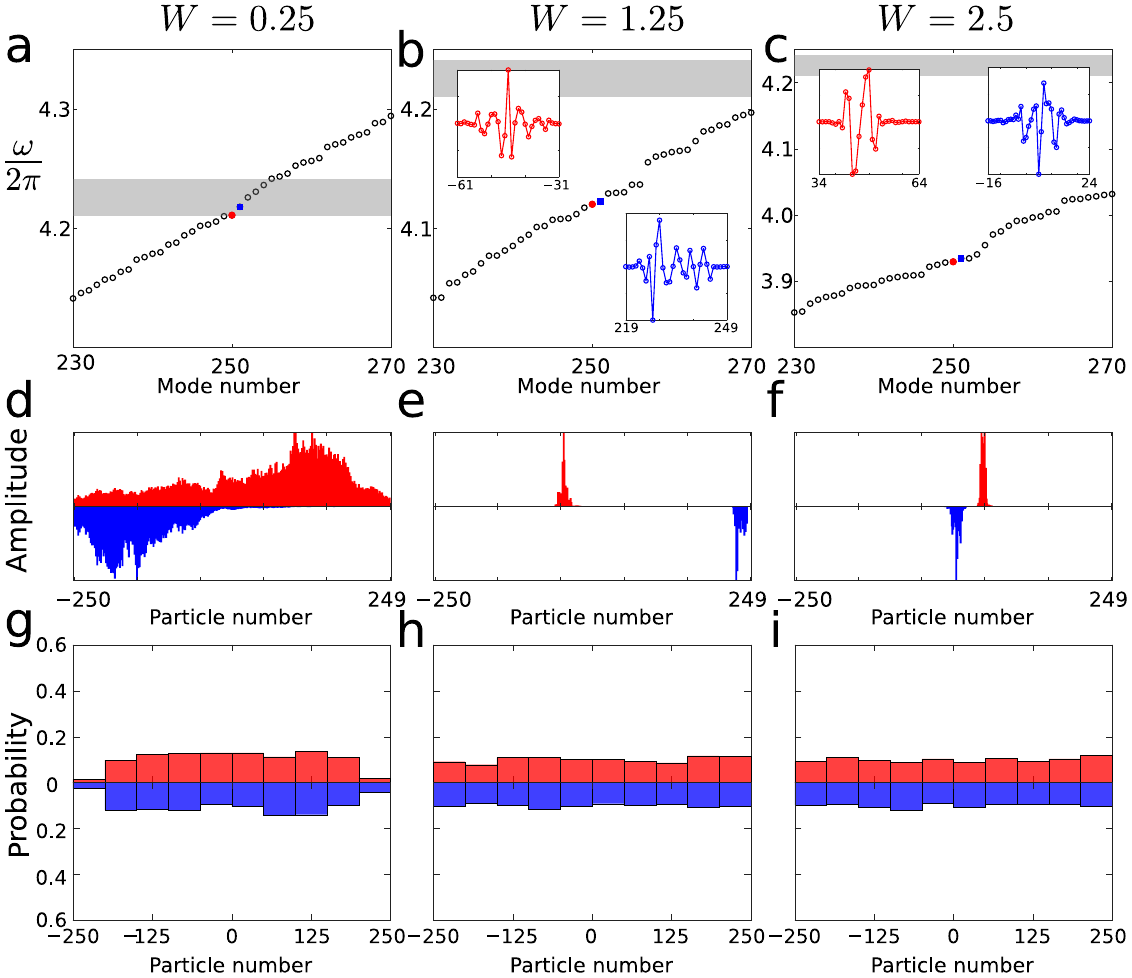}
\caption{Eigen-frequencies and two modes in the center of frequency spectrum of a finite spring-mass chain ($K_a=1.05$ and $K_e=1$) under \textit{random} disorder with disorder ratio $W=W_a=2W_e$ for (a) $W=0.25$, (b) $W=1.25$, and (c) $W=2.5$. These correspond to three typical realizations with different disorder strengths on the green curve in Fig.~\ref{FIG5}(c). Non-chiral nature of the two modes (marked) is shown in the insets.
(d)--(f) Spatial profile of these two modes for one disorder realization.
(g)-(h) Histogram of the Center of Mode (COM) obtained by $1000$ disorder realizations, describing the probability of COM lying in 50 particle interval throughout the chain.}
\label{FIG12}
\end{figure}

\def\bibsection{\section*{References}} 

\bibliographystyle{apsrev4-2}

\bibliography{TAI_references} 

\end{document}